# *Control of skyrmion chirality in Ta/FeCoB/TaO$_x$ trilayers by TaO$_x$ oxidation and FeCoB thickness*

R. Kumar[1], C.E. Fillion[1], B. Lovery[1], I. Benguettat-El Mokhtari[2], I. Joumard[1], S. Auffret[1], L. Ranno[3], Y. Roussigné[2], S.M.Chérif[2], A. Stashkevich[2], M. Belmeguenai[2], C. Baraduc[1], H. Béa*[1,4]

[1] Univ. Grenoble Alpes, CEA, CNRS, Spintec, 38000 Grenoble, France

[2] Univ. Sorbonne Paris Nord, LSPM, CNRS, UPR 3407, F-93430 Villetaneuse, France

[3]Univ. Grenoble Alpes, CNRS, Néel Institute, F-38042 Grenoble, France

[4] Institut Universitaire de France (IUF)

**Abstract:**

Skyrmions are magnetic bubbles with nontrivial topology envisioned as data bits for ultrafast and power efficient spintronic memory and logic devices. They may be stabilized in heavy metal/ferromagnetic/oxide trilayer systems. The skyrmion chirality is then determined by the sign of interfacial Dzyaloshinskii-Moriya interaction (DMI). Nevertheless, for apparently identical systems, there is some controversy about the DMI sign. Here we show that the degree of oxidation of the top interface and the thickness of the ferromagnetic layer play a major role. Using Brillouin light scattering measurements in Ta/FeCoB/TaO$_x$ trilayers, we have demonstrated a sign change of the DMI with the degree of oxidation of FeCoB/TaO$_x$ interface. Using polar magneto-optical Kerr effect microscopy, we consistently observed a reversal of the direction of current-induced motion of skyrmions with the oxidation level of TaO$_x$, attributed to their chirality reversal. In addition, a second chirality reversal is observed when changing the FeCoB thickness, probably due to the proximity of the two FeCoB interfaces in the ultrathin case. By properly tuning the chirality of the skyrmion, the spin transfer and spin orbit torques combine constructively to enhance the skyrmion velocity. These observations thus allow envisioning an optimization of the material parameters producing highly mobile skyrmions in this system and could be extended to other stacks. This chirality control also enables a more versatile manipulation of skyrmions and paves the way towards multidirectional devices.

## I. Introduction

Magnetic skyrmions are swirling chiral magnetic textures [1], which have recently attracted great attention since they are envisioned as data bits for next-generation ultra-fast and power efficient memory and logic devices [2], [3]. Initially, skyrmions were observed at low temperature in bulk magnetic materials lacking inversion symmetry such as MnSi [4], FeGe [5] and FeCoSi [6], [7]. More recently, skyrmions have been observed at room temperature in heavy metal/ferromagnet/oxide (HM/FM/$MO_x$) [8], [9], [10] and heavy metal/ferromagnet/heavy metal (HM/FM/HM) [9], [11], [12] trilayers. Their existence is favored by Dzyaloshinskii-Moriya interaction (DMI) [13], [14], which promotes non-collinear magnetic moments, and they are further stabilized by dipolar energy [15], [16]. In multilayered heterostructures with broken inversion symmetry, for instance in HM/FM/$MO_x$ and HM/FM/HM with different HMs, DMI has an interfacial origin. It is mediated by conduction electrons subjected to spin-orbit coupling at HM/FM interface (Fert-Levy type DMI) [17], [18] and to Rashba field at the FM/$MO_x$ interface (Rasbha type DMI) [19], [20].

The applicative potential of skyrmions is due to their micrometric to nanometric size and to their efficient manipulation with an injected electrical current. In HM/FM/$MO_x$ heterostructures, current-induced motion of skyrmion is mainly driven by spin-orbit torque (SOT), which results from spin Hall effect in the HM layer [9], [10], [21], [22], [23]. The direction of their motion is governed by two factors: (i) the sign of the spin Hall angle in the HM and (ii) the skyrmion chirality [9], [10], [22], [24], [25]. Chirality, defined by the sense of rotation of spins within the skyrmion domain wall, is determined by the sign of DMI [26]. The total DMI in such trilayer is usually considered as the sum of the DMI stemming from the two interfaces and is often called effective DMI. These two DMI terms are independent for thick enough ferromagnetic layer, but may become coupled when the ferromagnet thickness is below 2-3 monolayers [20], [27]. The same interface with inversion of the stacking order is expected to give the same magnitude of DMI, but with a change of sign. However, this is true only if the crystalline structure remains the same, and film growth direction often breaks this symmetry. For instance, for nearly zero initial DMI values in HM/FM/$MO_x$ samples, with the two DMI contributions compensating each other, $He^+$ ion irradiation induces a non-equivalent intermixing of top and bottom interfaces, which leads to a non-zero DMI [28].
In this type of HM/FM/$MO_x$ systems, different material variations have been used to achieve a modulation of the strength [29], [30], [31] and in some cases of the sign [32], [33] of effective DMI constant $D$: variation of the FM thickness [29], modification of the adjacent heavy-metal underlayer in HM/(FM=CoFeB,Co)/MgO heterostructures [32], [33], insertion of a Pt wedge layer in Ta/FeCoB/Pt(wedge)/MgO thin films [30], and the oxidation state of FM/$MO_x$ interface in Pt/Co/$MO_x$ [31]. Recently, it has been experimentally shown that the DMI sign in Cu/CoFe/$CoFeO_x$/Ta and Cu/CoFe/Ta(t)/$TaO_x$/Ta is changed when the oxide type (either $CoFeO_x$ or $TaO_x$) or Ta thickness is modified [34]. Besides, experimental results showed that the oxygen adsorbed content at Pd/Co/Ni/O surface can invert DMI sign and domain wall chirality [35]. A similar sign crossover of DMI has also been theoretically predicted in Ir/Fe/O depending on the O coverage [36]. Finally, recent experiments showing a non-zero DMI in nominally symmetric structures of Au/Fe/Au were attributed to different strains at the top and bottom interface [37].

Interestingly, a spread of DMI values and more importantly an inversion of DMI sign are obtained in the literature for nominally similar trilayers such as Ta/FeCoB/oxide stacks, where relatively small DMI values (several tens of µJ/m² typically) are generally found. The reported chirality for various stacks are summarized in Table *1*, together with the method of measurement. For instance, in similar Ta/FeCoB/MgO stacks, opposite DMI signs and domain wall chiralities are found by different teams [38], [39], [32], a variation of MgO thickness in Ta/FeCoB/MgO/Pt samples lead to an inversion of DMI sign [40] and the insertion of an ultrathin Ta layer at the FeCoB/MgO interface gives a right-handed chirality, as shown in the supplementary materials of [41]. Similarly, in Ta/FeCoB/TaOx stracks, either left- of right-handed chiralities have been observed [10], [42], [43], [44], [45] . This discrepancy in the sign of DMI in nominally similar sputter-deposited systems might be attributed to the small and nearly compensating DMI amplitude from top and bottom interfaces. The effective DMI, which is the sum of both contributions, might thus change sign due to different target compositions, in particular for FeCoB, or to different machines and deposition details used. For instance, the quality and oxidation level of the FeCoB/oxide interface might play an important role, depending on whether the oxide is deposited by RF sputtering or by DC sputtering of the metal followed by post-oxidation. The growth rate and sample roughness may also play a role. However, a clear picture of the DMI in such Ta/FeCoB/oxide systems is still missing.

| Oxide or overlayer | Chirality | Method | Reference |
|---|---|---|---|
| MgO | CW | FDWM | [38] |

| MgO | CW | CIM | [39] |
|---|---|---|---|
| MgO | CCW | CIM | [32] |
| MgO/Pt | CW / CCW | BLS | [40] |
| Ultrathin Ta/MgO | CW | CIM | [41] |
| $TaO_x$ | CCW | CIM | [10] [42] |
| $TaO_x$ | CW | BLS | [43] |
| $TaO_x$ | CW | BLS+CIM | [44] [45] |
| $TaO_x$ | CW / CCW | BLS+CIM | Here and [46] |

Table 1 Summary of the chirality obtained for Ta/FeCoB/oxide system with various oxides and overlayers. CW (CCW) stands for clockwise (resp. counter clockwise). FDWM stands for field driven domain wall creep motion, BLS stands for Brillouin Light scattering and CIM stands for current induced motion of skyrmions or domain walls.

Here, we report on the observation of two DMI sign crossovers in Ta/FeCoB/$TaO_x$ sputtered samples. One is driven by the FeCoB/$TaO_x$ top interface oxygen content and the other one by the variation of FeCoB thickness. These findings are deduced from Brillouin Light Spectroscopy (BLS) measurements combined with a detailed study of the direction of motion of skyrmions under DC current using polar magneto-optical Kerr Effect (p-MOKE) microscopy. These observations may explain the discrepancies observed in the literature about this system. Moreover, we show that tuning DMI sign and amplitude in such trilayer is a tool to optimize the current induced motion of skyrmions, by constructively combining the effects of spin transfer and spin orbit torques.

## II. Sample preparation

Our samples consist in substrate/Ta(3)/FeCoB(0.6-1.6)/Ta (0.5-1.0)$O_x$/Al (0.5)$O_x$ (nominal thicknesses of deposited metals in nm) stacks grown by DC magnetron sputtering deposition on a thermally oxidized Si/$SiO_2$ 100 mm-diameter wafer. The composition of the FeCoB target is $Fe_{72}Co_8B_{20}$. After top Ta deposition, a natural oxidation step (150 mbar oxygen pressure for 10 s) is performed to oxidize the Ta layer. Then a 0.5 nm cap layer of Al is sputter-deposited *in-situ* and oxidizes in air. Samples were annealed at 225°C under vacuum for 30 min. Finally, we deposited *ex-situ* a thicker oxide of $HfO_2$ by atomic layer deposition, to protect the stack from further oxidation.

Layers with uniform thickness were deposited with the substrate facing the target (on-axis deposition) whereas thickness gradients were obtained by shifting the sample with respect to the target (off-axis deposition). For BLS measurements, we have studied samples of uniform thickness with different thicknesses of the top-Ta, giving rise to various oxidation states at the FeCoB/$TaO_x$ interface, namely three types of interfaces: FeCoB/$FeCoBO_x$/$TaO_x$ for overoxidized interface, FeCoB/$TaO_x$ for optimally oxidized interface and FeCoB/Ta/$TaO_x$ for underoxidized interface. For p-MOKE measurements, the sample used was a double-wedge sample, as described in [45]: the thickness gradient of FeCoB, along the vector $\hat{x}$ is perpendicular to the following wedge of Ta, along the vector $\hat{y}$. This second wedge leads to a decreasing degree of oxidation (x) of $TaO_x$ layer when Ta thickness increases. The exact values of the oxidation degree and of the oxide thickness are unknown. This double wedge sample is particularly well suited to investigate the effects of FeCoB thickness and of $TaO_x$ oxidation state on DMI, since this combinatorial approach avoids sample-to-sample fluctuations that could hide small variations in the DMI amplitude.

## III. DMI sign crossover measured by BLS

In order to explore the effect of the oxidation state of FeCoB/$TaO_x$ interface on DMI energy, we have performed BLS measurements on samples with uniform FeCoB thickness, with different thicknesses of top Ta. A magnetic field was applied in the film plane, perpendicular to the laser incidence plane, to probe spin waves propagating in the plane direction perpendicular to the applied magnetic field, in the Damon−Eshbach (DE) geometry. The applied magnetic field is above the saturation field of the sample, as determined from magnetometry loops. The frequency shift $\Delta f$ between Stokes ($f_S$) and anti-Stokes ($f_{AS}$) frequency lines is then deduced. The frequency offset due to the set-up was evaluated from surface acoustic line positions. The spin-wave vector is given by $k_{sw} = 4\pi \sin(\theta_{inc})/\lambda$ , where $\theta_{inc}$ (between 10 and 60°) is the angle of incidence and $\lambda = 532$ nm is the wavelength of the incident laser. Spectra for top Ta thicknesses of 0.6, 0.7 and 0.8nm, represented in Figure *1*(a), show both Stokes and anti-Stokes frequencies and the corresponding Lorentzian fits. Figure *1*(b) represents the fits of both resonances reported in the positive frequency range to illustrate the frequency difference $\Delta f$ between Stokes and anti-Stokes frequency lines $\Delta f = |f_S| - |f_{AS}|$. $\Delta f$ is linked to the DMI constant $D$ by $\Delta f = \frac{2\gamma}{\pi M_S} k_{SW} D$, where $\gamma$ is the gyromagnetic ratio and $M_s$ the saturation magnetization of the sample. The DMI constant deduced from the measured $\Delta f$ as a function of top Ta thickness is shown in Figure *1*(c). Our convention, similar to reference [44], is that positive (resp. negative) DMI corresponds to right-handed or clockwise (resp. left-handed or counter-clockwise) chirality. For the thinner Ta region ($t_{top\text{-}Ta}$ = 0.6 nm), which thus corresponds to over-/optimally oxidized FeCoB/$TaO_x$ interface, $\Delta f$ is positive, which results in a positive effective DMI $(D > 0)$. For thicker Ta ($t_{top\text{-}Ta} \geq 0.7$ nm), for optimally-/underoxidized FeCoB/$TaO_x$ interface, $\Delta f$ changes sign and becomes negative which results in a negative effective DMI $(D < 0)$.

The measured (effective) DMI results from the contribution of the bottom Ta/FeCoB interface, which is the same for all samples, and of the top interface that may change with oxidation state. From the literature, the bottom Ta/FeCoB interface gives a small Fert-Levy type DMI [17], which surface value $D_s = Dt$, where $t$ is the ferromagnet thickness, is of the order of 20-30 fJ/m [28], [33]. In order to deduce the contribution from top FeCoB/$TaO_x$ interface in our system, we used a value of 30 fJ/m for the bottom Ta/FeCoB interface, taking an upper limit, which remains small with respect to the total DMI value and to the contribution from the top interface. The considered ferromagnetic thickness was 0.55 nm due to the dead layer of 0.6 nm extracted by magnetometry measurements along the FeCoB wedge, using the extrapolation at zero magnetization of the moment versus FeCoB thickness [47] (not shown). Figure *1*(c) shows the estimated values of DMI for each interface: we observe that the DMI from the top interface changes sign with the oxidation state.

Theory has predicted that DMI depends on the oxidation state at the ferromagnet/oxide interface due to hybridization of transition metal *3d* orbitals (Fe and Co) with oxygen *2p* orbitals [36]. From our observations, DMI is positive for high/optimum oxidation state of FeCoB/$TaO_x$ interface (thin Ta layer), and becomes negative for a lower oxidation state (thicker Ta). For optimally/overoxidized interface, the top interface, FeCoB/$TaO_x$, is expected to possess Rashba type DMI [19], [20], [31], [44], which is related to the interfacial electric field. Hence, it depends mostly on the oxidation state and charge transfer at FeCoB/$TaO_x$ interface [20], [34], [36]. We thus conclude that DMI of Rashba origin gives a positive DMI contribution at the FeCoB/$TaO_x$ interface. By contrast, the underoxidized FeCoB/Ta/$TaO_x$ interface shows a negative DMI. This observation is consistent with the presence of metallic Ta at the top interface. Fert-Levy mechanism is known to lead to positive DMI in Ta/FeCoB [28], [33], thus to negative DMI for the reversed stacking order. However, the effective negative DMI means that the absolute value is larger for the top FeCoB/Ta/$TaO_x$ interface. This might be due to an intermixed interface resulting from the deposition of the heavy metal on top of FeCoB [28], to non-uniform interface likely composed of FeCoB/Ta and FeCoB/$TaO_x$ and/or to different strains at the top and bottom interfaces [37]. BLS measurements in Pt/Co/$TaO_x$ system give the same trend [48], but the curve is shifted by the large negative value of DMI from the bottom Pt/Co interface, typically of the order of -1.5 to -2 mJ/m$^2$ ($D_s$ = -1.3 to -2.2 pJ/m) [8], [49], [50].

This DMI sign crossover is similar to the one reported by Arora et al. [34], except that signs are opposite, which might be explained by the different samples. In their study, there is no heavy metal underlayer, the ferromagnet is Co rich ($Co_{90}Fe_{10}$) with no Boron and with a (111) texture due to Cu underlayer and the samples are not annealed. By contrast, in our present study, the ferromagnet is Fe rich with presence of Boron ($Fe_{72}Co_8B_{20}$), with Ta underlayer and with a bcc structure obtained after annealing. Smoother interfaces produced by annealing and migration of boron to the bottom Ta layer [48] (or possibly to the top Ta interface when underoxidized) might thus be critical parameters for DMI.

## IV. Magnetic characterization of Ta/FeCoB/TaOx double wedge

After measuring DMI for different oxidation states of FeCoB/$TaO_x$ interface in the previous section, we characterized the magnetic behavior of our double wedge sample, and the behavior of skyrmions under current. Polar MOKE measurements have been performed on double wedge samples at room temperature. A p-MOKE magnetometer (NanoMOKE®3 by Durham Magneto Optics Ltd.) was used to measure the hysteresis loops on the double wedge sample, every 2 mm throughout 100 mm diameter sample and a mapping of the magnetization at remanence (in % of saturation magnetization) was generated (See Figure *2*(a)). This remanence map measured under magnetic field perpendicular to the layer plane of the double wedge sample shows different magnetic behaviors depending on material thicknesses, as explained in our previous study [45]. Perpendicular magnetic anisotropy (PMA) region is characterized by a full remanence, in-plane anisotropy (IPA) region gives a hard-axis behavior with no remanence and paramagnetic (PM) state corresponds to non-magnetic (noisy) signal. The transition from PM to PMA to IPA for increasing FeCoB thickness is usual: for low FeCoB thicknesses, the dead layer explains the PM state at room temperature. For increasing FeCoB thickness, the effective anisotropy changes sign from positive, dominated by surface contribution in the PMA region, to negative, dominated by the dipolar contribution in the IPA region. The effect of the oxidation gradient of $TaO_x$ is less usual. The transition from PMA to PM for decreasing top Ta thickness is related to the increase of dead layer due to partial oxidation of FeCoB. The transition from PMA to PM for increasing top Ta thickness is explained by underoxidation. The direct contact of the ferromagnetic layer with metallic Ta is expected to reduce the magnetic moments on iron, as shown by *ab-initio* calculations [51], [52]. Thus, both overoxidation and underoxidation result in increasing the dead layer at room temperature. Here we consider the room temperature dead layer indistinctly due to a possible 'true' dead layer at zero temperature and to the loss of magnetic ordering for thin ferromagnetic thickness at finite temperature [47], [53], [45].

At room temperature and under zero applied magnetic field, thermally activated demagnetization leads to the formation of high-density stripe domains in regions close to transition between PMA and IPA regions and between PMA and PM regions [44]. These regions, where stripes form, are close to the white transition line surrounding the PMA region in Figure *2*(a). Moreover, on application of a small out-of-plane magnetic field (40 to 400 μT depending on location), the stripes transform into micron size skyrmions [45]. In the present paper, skyrmions were obtained by applying a magnetic field sweep: after magnetization saturation under a large out-of-plane field, the field is decreased to a small value with the same polarity.

## V. Two skyrmion chirality reversals

In order to confirm the inversion of DMI sign with the oxidation of $FeCoB/TaO_x$ interface and its influence on skyrmion chirality, we have studied the current induced motion of skyrmions on the double wedge sample at several locations, represented by numbers in Figure *2*(a). The two Néel skyrmion chiralities are depicted in Figure *2*(b). The DC current was locally injected using micro-wires bounded on top of the stack and a p-MOKE microscope was used for observing skyrmion motion under the resulting current density (see Supplementary videos SV1 to 8). The DC current sets the skyrmions in motion in a direction depending on their chirality: since Ta has a negative spin-Hall angle, left-handed or CCW (resp. right-handed or CW) Néel skyrmions are expected to move along (resp. against) the electron flow [24]. At large skyrmion velocity (typically m/s), a transverse motion of skyrmions adds to the longitudinal one, leading to skyrmion Hall effect [54], [55]. In the present case, this effect is not observed due to the low current density (~ $10^9$ A/m$^2$) [54]. In the following, we will use the direction of motion of skyrmions under current to deduce the skyrmion chirality and DMI sign.

First, we observed close to the PMA to IPA transition, for ~1.2 nm thick FeCoB, a change of skyrmion motion direction depending on the degree of oxidation. Skyrmions move along the current (opposite to the electron flow) in the thinner Ta region (location #7) and in the opposite direction for thicker Ta (location #6) (see Supplementary videos SV7, 8 for skyrmion motion in locations #6 and 7 respectively). This observation is in agreement with the DMI sign crossover observed by BLS in Figure *1*. In between these regions, highlighted by the white dashed line in Figure *2*(a), we observe nearly no motion of skyrmions under current, which we attribute to zero DMI. More precisely, for thin Ta (#7), skyrmions move along the injected current direction but, due to the proximity to the $D \approx 0$ region, their velocity is small. This direction of motion is consistent with our previous studies, which focused on the same material stack with even thinner Ta (CW chirality) and showed relatively high pinning [45].

In the PMA to PM transition for relatively thin FeCoB layers, equivalent to location #1, we previously measured right-handed CW chirality skyrmions, combined with positive DMI coefficient measured by BLS [44]. These findings contrast with the left-handed/CCW chirality discussed in the previous paragraph for thicker FeCoB layer (location #6) and might seem surprising. In fact, a second chirality reversal occurs along the transition between PMA and PM when FeCoB thickness varies for nearly constant oxidation state of FeCoB/$TaO_x$ interface. This is represented in Figure *2*(a) by the numbers #1 to 5. For thinner FeCoB (locations #2 and 1, the latter being close to the region studied in [45]), skyrmions move along the current i.e. opposite to the electron flow (see snapshots for location #2 in Figure 3(a-c) and videos SV1 and 2). By contrast, for thicker FeCoB (locations #4 and 5), skyrmions move along the electron flow and opposite to the injected current direction (see snapshots for location #4 in Figure 3 (d-f) and videos SV3 and 4). This inversion of direction is a signature of a change of skyrmion chirality from right-handed/CW for thin FeCoB to left-handed/CCW for thicker FeCoB. It is thus expected to be driven by a second DMI sign crossover. The negative DMI measured by BLS for thick top Ta and thick FeCoB (see Figure *1*) together with the measurements of positive DMI in the thin FeCoB region [44] also validates this DMI sign crossover with FeCoB thickness. We note here that the same DMI sign inversion with FeCoB thickness was obtained for a single FeCoB wedge sample, ie. for a constant oxidation state of the FeCoB/$TaO_x$ interface (not shown).

Interestingly, at the boundary between regions of positive and negative DMI, corresponding to location #3 whereDMI is expected to be zero , we observed a high density of bubbles that hardly move under current (see inset of Figure 3(g) and Supplementary videos SV5-6). They just show usual Brownian motion, expected for 300K experiments. In the case of zero DMI, skyrmions with Bloch domain wall may be stabilized by dipolar energy. The driving force on these Bloch skyrmions is perpendicular to the current, either in one direction or in the other depending on their chirality. As they have no preferred chirality, we interpret the absence of motion as the effect of repulsive interactions between closely packed bubbles that are driven in opposite directions.

To summarize these measurements on the double wedge sample, we used colored numbers in Figure *2*(a): red (resp. blue) corresponds to right-handed or CW (resp. left-handed or CCW) chirality and to a positive (resp. negative) DMI sign. The inversion of skyrmion motion direction, and thus of DMI sign, occurs at two positions highlighted in Figure *2*(a) by dashed white lines: one along the FeCoB wedge ($t_{FeCoB}$ ~ 1 nm, and $t_{top\text{-}Ta}$ ~ 0.92 nm) and another one along the $TaO_x$ wedge ($t_{top\text{-}Ta}$ ~ 0.87 nm, and $t_{FeCoB}$ ~ 1.2 nm). The thickness values given here are only the nominal thicknesses of the layers, since the real deposited thickness may vary from one deposition to another by 5-10%. This could be one reason why the nominal thickness corresponding to DMI crossover is different when extracted from BLS measurements (uniform samples) and from skyrmions motion (double wedge sample). Another reason could be that depositing samples of uniform thickness, with a rotation of substrate during deposition to get a homogeneous layer, and samples with thickness wedge, without substrate rotation, could lead to different roughness and in-plane anisotropies.

The mechanism for the chirality reversal and DMI sign crossover when varying FeCoB thickness might be due to the relatively large dead layer (around 0.6 nm [46]) related to the underoxidization of FeCoB/$TaO_x$ interface [56], [45]. Taking into account the dead layers, we may consider that the nominal thickness of 1 nm corresponds to an effective thickness of the ferromagnet significantly lower, of about 0.4 nm, i.e. around 3 monolayers. We may thus ascribe the observed variation of DMI sign with FeCoB thickness to the onset of a coupling between the two Ta/FeCoB and FeCoB/$TaO_x$ interfaces or to oscillatory behavior of $D_s$ for such thin films [27], [20]. Besides, we may also consider the influence of a change in band structure or a stronger contribution of interfacial disorder due to these ultrasmall thicknesses.

In addition, we note that the DMI sign inversions occur with a thickness variation of about 0.1 nm of FeCoB along the FeCoB wedge, (see locations #2 and #4 in Figure *2*a) and about 0.05 nm of Ta along the oxidation gradient (see locations #6 and #7 in Figure *2*a), ie. less than a monolayer in both cases. The origin of these DMI sign crossovers is probably linked to subtle material detail. We may wonder whether the growth rate may play a role since it varies by roughly 10% between locations #2 and #4 and between locations #6 and #7. However, we must rule out this hypothesis since we have also observed these transitions of DMI sign with both oxidation of $TaO_x$ and thickness of FeCoB for another sample with a smaller growth rate of about 10% (not shown).

The existence of these two DMI sign crossovers, with oxidation of FeCoB/$TaO_x$ interface or with FeCoB thickness might explain the discrepancies previously observed in the literature for the DMI sign in this stack.

## VI. Importance of skyrmion chirality control

Finally, we have studied in more details the influence of DMI sign on skyrmion velocity along the FeCoB wedge that presents the unexpected chirality reversal. The current lines are radial close to the point contact, as can be observed by the motion of skyrmions in the videos SV1-2, SV5-6. Using current conservation, we may thus estimate a current density of ~1.3 $10^9$ A/m$^2$ for 5 mA applied current at a distance of 150 µm ($\frac{5mA}{4nm\times 2\pi\times 150\mu m}$), distance at which skyrmion velocity was measured. Figure 3(g) represents the evolution of skyrmion velocity for 5 mA injected current in the positive DMI location (#1-2), in the DMI sign crossover location (#3) and in the negative DMI location (#4-5). With our convention, a positive velocity corresponds to a motion along the injected current direction, thus to a right-handed/CW chirality. When crossing the boundary where DMI changes sign and skyrmion chirality is inverted, the sign of skyrmion velocity is inverted as well (see also Supplemental Material videos SV1-6 for p-MOKE videos of these experiments).

Interestingly, as shown in Figure 3(g), the typical velocities at locations #5 ($D < 0$) is around -53 *µm/s*, whereas it is only about +10 *µm/s* at location #1 ($D > 0$), similar to the results of other teams on this system [10], for equivalent current densities. Besides, the left-handed/CCW chirality skyrmions ($D < 0$ region) can be moved with a current as small as 0.5 mA, which confirms a significantly larger mobility and lower pinning of these skyrmions with left-handed/CCW chirality. In this case, the spin orbit torque (SOT) favors a motion along the electron flow, similarly to Zhang and Li spin transfer torque (STT) [57]. To estimate their relative contributions, we calculate the current distribution within the stack. By using the tabulated resistivities of Ta, Fe and Co ($R_{/\square}(Ta)\approx 44\ \Omega/_{\square}$ and $R_{/\square}(Fe_{0.9}Co_{0.1})\approx 95\ \Omega/_{\square}$) and the thicknesses of our sample, we roughly estimate that one third of the current flows through the FeCoB and two thirds in the Ta underlayer. Spin transfer torque within the FeCoB might thus play a non-negligible role in skyrmion motion. Hence, in the left-handed/CCW case, both torques act together to accelerate skyrmions. By contrast, if the chirality is inverted, STT is now opposite to SOT, which slows down skyrmion motion. This addition of STT for left-handed/CCW skyrmions might thus explain that, in this negative DMI region, skyrmions can be moved with a ten times lower current as compared to the positive DMI region. In addition to the amplifying or reducing effect of STT, a possible variation in pinning strength (due to the slightly different growth conditions along the wedges) may also contribute to the difference in skyrmion speed. As a result, the left-handed/CCW skyrmions with large mobility and low pinning are the most promising for applications.

## VII. Conclusion

In this paper, we have shown the control of DMI sign and skyrmion chirality using material parameters, such as ferromagnetic thickness and/or oxidation of top FeCoB/$TaO_x$ interface. The DMI sign inversion along the oxidation gradient is understood in terms of a change of mechanism, from Fert-Levy to Rashba when increasing the oxidation level. In the intermediate region where DMI is zero, we observed no current induced motion, which is likely due to repulsive interactions between Bloch bubbles of random chirality. Skyrmions with counterclockwise chirality show higher mobility under current than skyrmions with clockwise chirality, which we attribute to the additive combination of spin transfer torque in the FeCoB and spin orbit torque from the Ta heavy metal underlayer. Moreover, obtaining a DMI crossover is essential for an efficient DMI control by gate voltage since electric field effect can be much larger close to a transition. As oxidation is a key parameter for DMI sign in this system, ion migration controlled by a gate voltage [58] can thus dynamically switch DMI and skyrmion chirality [46]. The control of chirality enables a tuning of the skyrmions speed and trajectory, which may lead to novel functionalities in skyrmion based logic, memory and neuromorphic devices.

Fig. 1

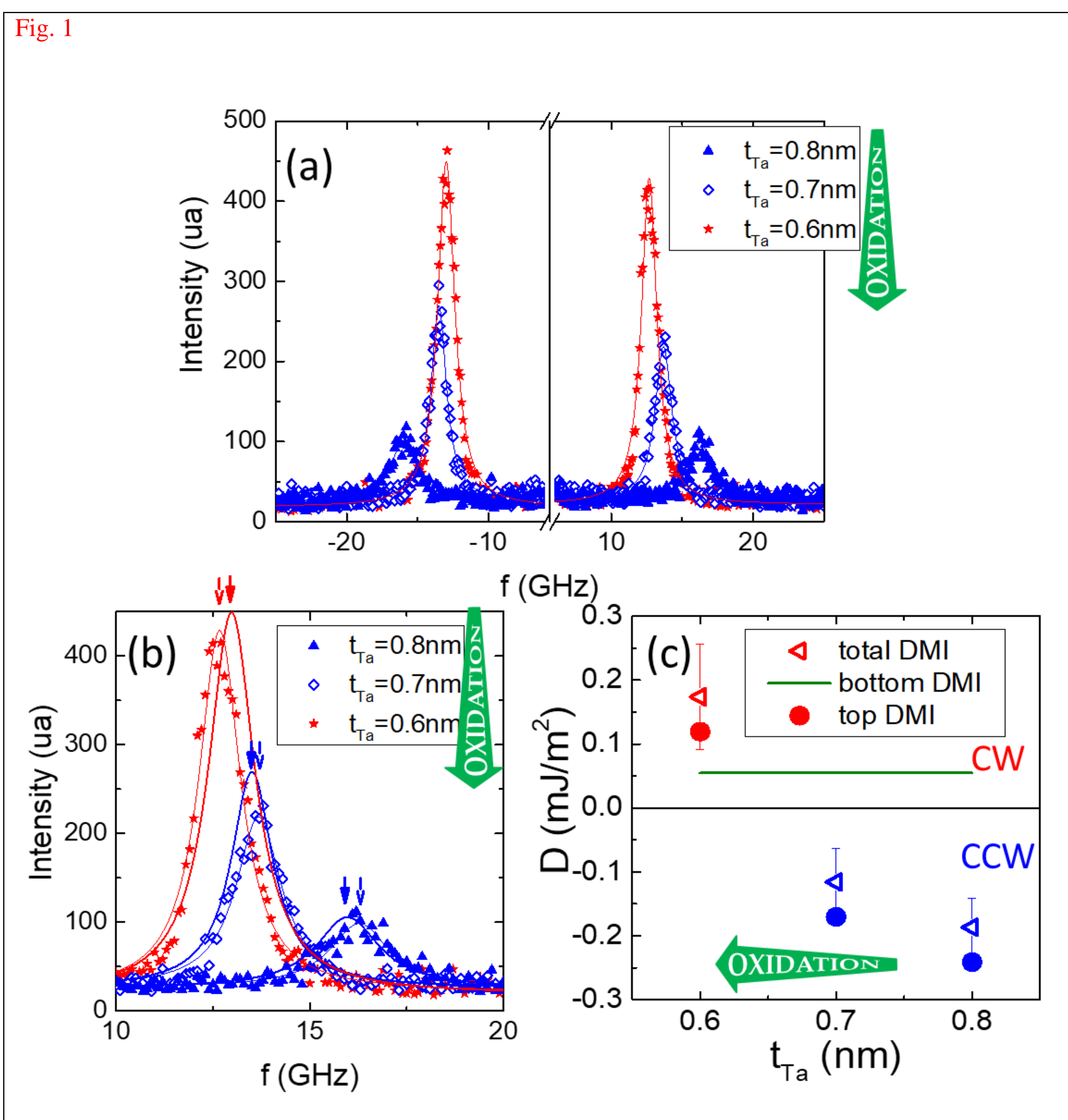

Figure 1. (a) BLS spectra (symbols) for Ta/FeCoB(1.15 nm)/TaOx samples for varying top Ta thickness taken at $k_{sw}$=20.45 $\mu m^{-1}$. The solid lines are Lorentzian fits to the spectra. (b) Superimposition of the Stokes and anti-Stokes lines in the positive frequency region in order to better visualize their frequency difference. The thin solid lines are the Lorentzian fits of anti-Stokes resonance and the thick solid lines are the fits of the Stokes line. The position of both peaks are highlighted by thin or thick arrow

respectively. (c) DMI value extracted from (a), and its variation as a function of top Ta thickness. The DMI value of the bottom interface is taken from literature, allowing to deduce the top interface DMI and its variation.

Fig. 2

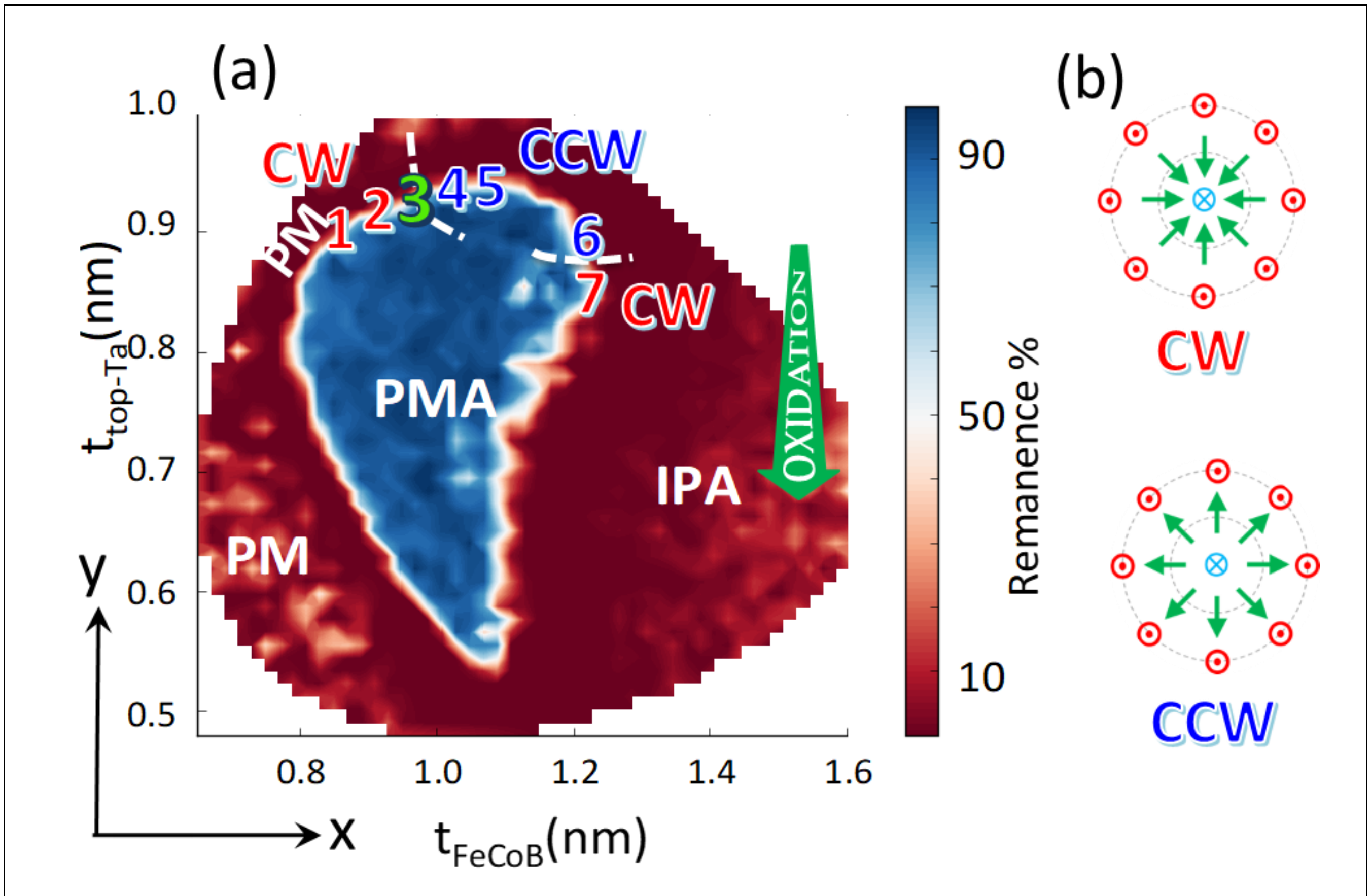

Figure 2 (a) Perpendicular remanence map of the Ta(3)/FeCoB(0.6-1.6)/Ta(0.5-1)Ox/$AlO_x$(0.5) double wedge generated by performing p-MOKE hysteresis loops every 2 mm. Perpendicular magnetic anisotropy (PMA), in-plane anisotropy (IPA) and paramagnetic (PM) phases are indicated. Numbers show locations where skyrmion motion has been studied by p-MOKE microscopy: in red (resp. blue) CW (resp. CCW) chirality was observed. At the boundaries between CW and CCW regions, indicated by white dashed lines and a green number, no motion is observed under current. (b) Schematic representation of skyrmions of clockwise (CW, top) or counterclockwise (CCW, bottom) chirality.

Fig. 3

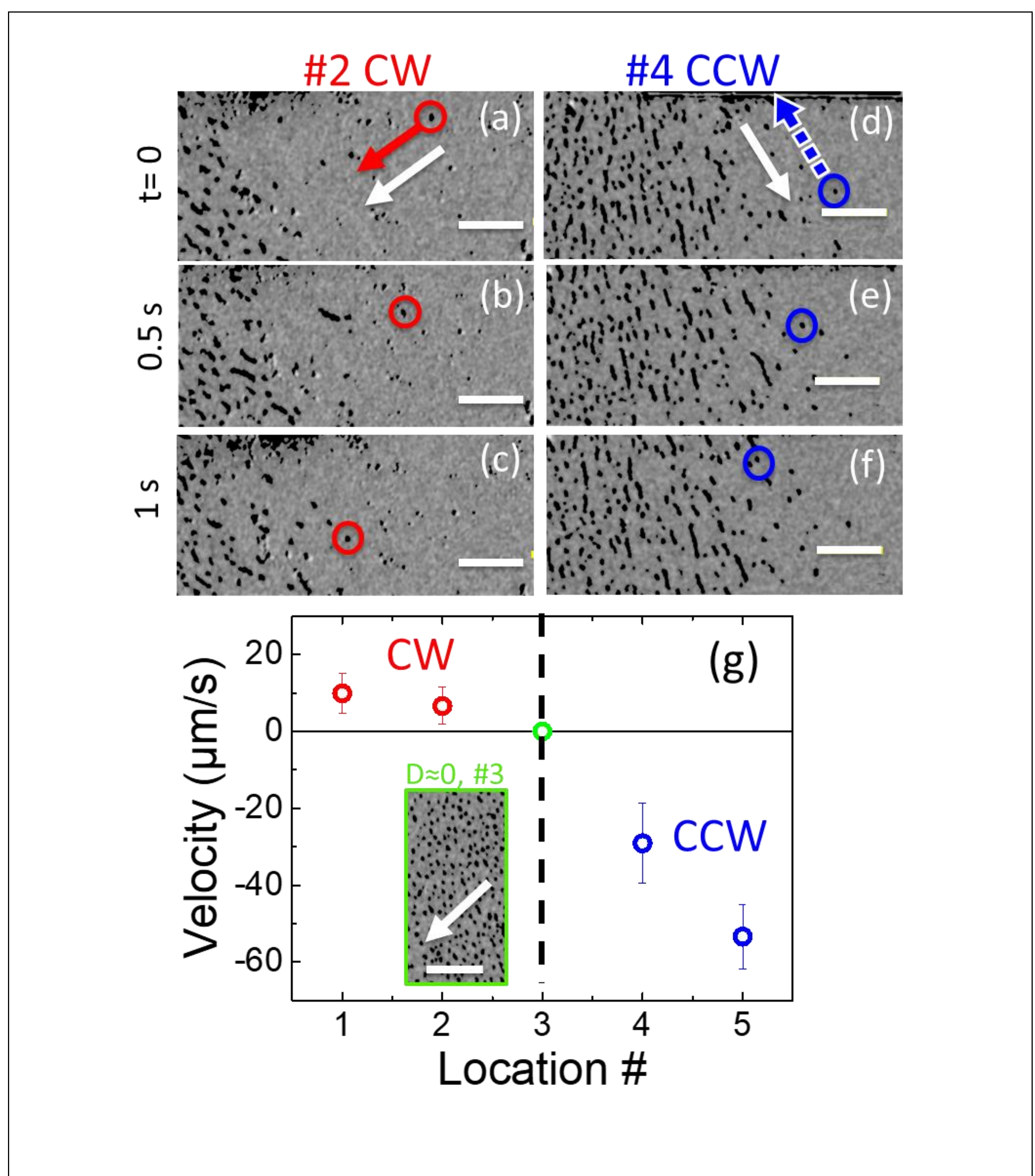

Figure 3 (a-f) p-MOKE microscopy images taken under 400 µT perpendicular field while applying a DC current of +5 mA and captured at 0 (a,d), 0.5 (b,e) and 1 s (c,f) time intervals, demonstrating current induced motion of skyrmions: (a-c) in location #2, skyrmions move along the injected current direction, consistent with CW chirality (see videos SV1-2), (d-e) in location #4, skyrmions move opposite to the injected current direction, consistent with CCW chirality (see videos SV3-4). The direction of injected current is shown by the white arrows and the white bar corresponds to 25 µm. The circles and arrows highlight the typical behavior of skyrmions. Locations #2, #3 and #4 are the same as in Fig. 2. (g) Skyrmion velocity under 5 mA corresponding to a current density of 1.3 $10^9$ A/$m^2$, measured at locations #1 to 5. A positive velocity corresponds to a motion of skyrmions along the injected current direction. The error bars mainly come from the statistical dispersion of velocities measured on several skyrmions (typically five). Inset: p-MOKE microscopy image in location #3 of the DMI sign crossover ($D \approx 0$), no detectable displacement of bubbles is observed under the same injected current, except usual Brownian motion (see videos SV5-6).

## ASSOCIATED CONTENT:

Supporting Information

## AUTHOR INFORMATION:

Corresponding Author

*E-mail: helene.bea@cea.fr.

ORCID

Raj Kumar: 0000-0003-4001-0011

Charles-Elie Fillion : 0000-0002-8440-0401

Bertrand Lovery: 0000-0002-3671-0608

Laurent Ranno: 0000-0002-1901-2285

Salim-Mourad Chérif: 0000-0003-4350-9379

Mohamed Belmeguenai: 0000-0002-2395-1146

Claire Baraduc 0000-0001-7592-0993

Hélène Béa: 0000-0002-3762-4795

**Notes**

The authors declare no competing financial interest.

## ACKNOWLEDGMENTS

The authors thank A. Fassatoui, N. Chaix and G. Gay for their help during the process for oxide deposition and pads deposition/fabrication and M. Chshiev for fruitful discussions. The authors acknowledge funding by the French ANR (contract ELECSPIN n°ANR-16-CE24-0018, contract ADMIS n°ANR-19-CE24-0019), Nanosciences Foundation and by the People Programme (Marie Curie Actions) of the European Union's Seventh Framework Programme (FP7/2007-2013) under REA grant agreement n. PCOFUND-GA-2013-609102, through the PRESTIGE programme coordinated by Campus France.

## VIDEOS OF P-MOKE MICROSCOPE IMAGES UNDER DC CURRENT

p-MOKE microscopy videos taken under current application via the microbonded wire. Pads (rectangles of 100 by 800 µm) have been deposited atop as reference locations on the sample. The white bars in SV1-6 are 100 µm. Skyrmions are black round shape dots that sometimes elongate as stripes.

**Video SV1.mp4** taken at location #2 under 400 µT perpendicular magnetic field and +5 mA current, corresponding to Figure 3(a-c)

**Video SV2.mp4** taken in location #2 under 400 µT perpendicular magnetic field and -5 mA current

**Video SV3.mp4** taken in location #4 under 400 µT perpendicular magnetic field and +5 mA current, corresponding to Figure 3(d-f)

**Video SV4.mp4** taken in location #4 under 400 µT perpendicular magnetic field and -5 mA current

**Video SV5.mp4** taken in location #3 under 400 µT perpendicular magnetic field and +5 mA current, corresponding to inset of Figure 3(g)

**Video SV6.mp4** taken in location #3 under 400 µT perpendicular magnetic field and -5 mA current

**Video SV7.mp4** taken in location #6 under 400 µT perpendicular magnetic field and -40 mA current (ie. going upwards). The image size corresponds to 118 by 73 µm².

**Video SV8.mp4** taken in location #7 under 400 µT perpendicular magnetic field and -40 mA current (ie. going upwards). The image size corresponds to 126 by 99 µm².